\documentclass[aps,prb,groupedaddress]{revtex4}
\usepackage[dvips]{graphicx}

\begin{document}




\title{Microscopic theory of single-electron tunneling through 
molecular-assembled metallic nanoparticles }

\author{ Yongqiang Xue $^{*}$ and Mark A. Ratner}
\affiliation{Department of Chemistry and Materials Research Center, 
Northwestern University, Evanston, IL 60208}
\date{\today}

\begin{abstract}
We present a microscopic theory of single-electron tunneling through metallic 
nanoparticles connected to the electrodes through molecular bridges. It  
combines the theory of electron transport through molecular junctions 
with the description of the charging dynamics on the nanoparticles. We 
apply the theory to study single-electron tunneling through a gold 
nanoparticle connected to the gold electrodes through two representative 
benzene-based molecules. We calculate the background charge on the 
nanoparticle induced by the charge transfer between the nanoparticle and 
linker molecules, the capacitance and resistance of molecular junction using 
a first-principles based Non-Equilibrium Green's Function theory. We 
demonstrate the variety of transport characteristics that can be achieved 
through ``engineering'' of the metal-molecule interaction. 
\end{abstract}


\maketitle


Chemically tailored metallic and semiconducting nanoparticles and their 
assemblies have become the model system for studying the fundamental 
physics and chemistry of nanostructured materials.~\cite{Heath98} 
The advancement of the fabrication process using self-assembly~\cite{Heath98} 
or biodirected-assembly technique~\cite{Bio} with molecular 
recognition has made devices based on single and assembled nanoparticles 
attractive candidates in applications including 
single-electronics, novel biosensors and nanophotonic 
devices.~\cite{Heath98,Bio,Andres96,SE} For single-electron 
device applications, the possibility of accurate control over particle 
size and density provides a substantial advantage over conventional 
granular metal thin films and suggests possible room-temperature 
operation.~\cite{Heath98,Andres96} Understanding electrical transport in 
such metal-molecule composite systems is therefore an important problem. 

In the ``orthodox'' theory of single-electron tunneling through a metallic 
island separated from the electrodes by insulating gap,~\cite{IN92} 
the physics of Coulomb blockade can be well understood 
using simple circuit-level theory, with circuit parameters including 
junction resistance, capacitance and charging energy often obtained 
by fitting experimental data and/or using simple electrostatic 
considerations. In contrast, for the nanoparticle-based single-electron 
devices, the transfer of single 
electron is achieved by tunneling through the linker molecule connecting 
the nanoparticles to the electrodes and to each other.~\cite{Andres96,SE} 
The linker molecules can be coupled strongly to the electrodes and 
nanoparticles through appropriate end groups, and the tunneling barrier is 
induced by the energy mismatch of the molecular levels relative to the metal 
Fermi-level.~\cite{JGA00} For the molecular junction, the metallic 
screening of the applied electric field occurs over a distance comparable to 
the size of the molecule~\cite{Xue01,Xue03}. Consequently, the capacitance 
of the molecular junction becomes an electro-chemical quantity rather than 
a geometrical quantity, whose evaluation requires a self-consistent 
analysis of the charge/potential response 
within the molecular junctions.~\cite{Buttiker93} 
In addition, charge transfer between the linker molecule and the metals will 
lead to fractional charge on the nanoparticle,~\cite{Xue01,Xue03} 
giving rise to an intrinsic background charge (even in the absence of 
charged impurities and gate voltages), which varies in general 
with the applied bias voltage and may affect significantly the 
current-voltage characteristics.~\cite{Tinkham91}
Better understanding of such molecular-assembled single-electron devices 
therefore requires combining the single-electron tunneling effect with 
a microscopic description of the electronic processes in the molecular 
tunnel junctions. 

In this paper we describe a microscopic theory of single-electron tunneling 
through molecular-assembled metallic nanoparticles by combining a 
first-principles Non-Equilibrium Green's Function (NEGF) theory of molecular 
transport~\cite{Xue01,Xue03} with a real-time perturbation theory of 
the reduced density matrix describing the charging dynamics of the 
metallic nanoparticles.~\cite{Schon94,Schon98} In the lowest-order 
perturbation theory in the reduced conductance of the molecular junction 
$\alpha_{0}=\frac{h}{2e^{2}R_{T}}$ ($R_{T}$ is the junction resistance), the 
theory formally reduces to the rate equation of the ``orthodox'' theory of 
Coulomb blockade,~\cite{IN92} but allows us to take full account of 
metal-molecule interaction at the atomic scale. In particular, the background 
charge, capacitance and tunneling rate can all be 
obtained from the first-principles theory of the molecular tunnel junction. 
We apply the theory to single-electron tunneling through gold nanoparticles 
connected to the gold electrodes through two benzene-based molecules-biphenyl 
dithiolate (BPD) and difluorobenzene (FBF) molecules and show the the 
variety of current/conductance-voltage characteristics that can obtained 
through ``engineering'' of the metal-molecule interaction. 

\begin{figure}
\includegraphics[height=3.0in,width=3.0in]{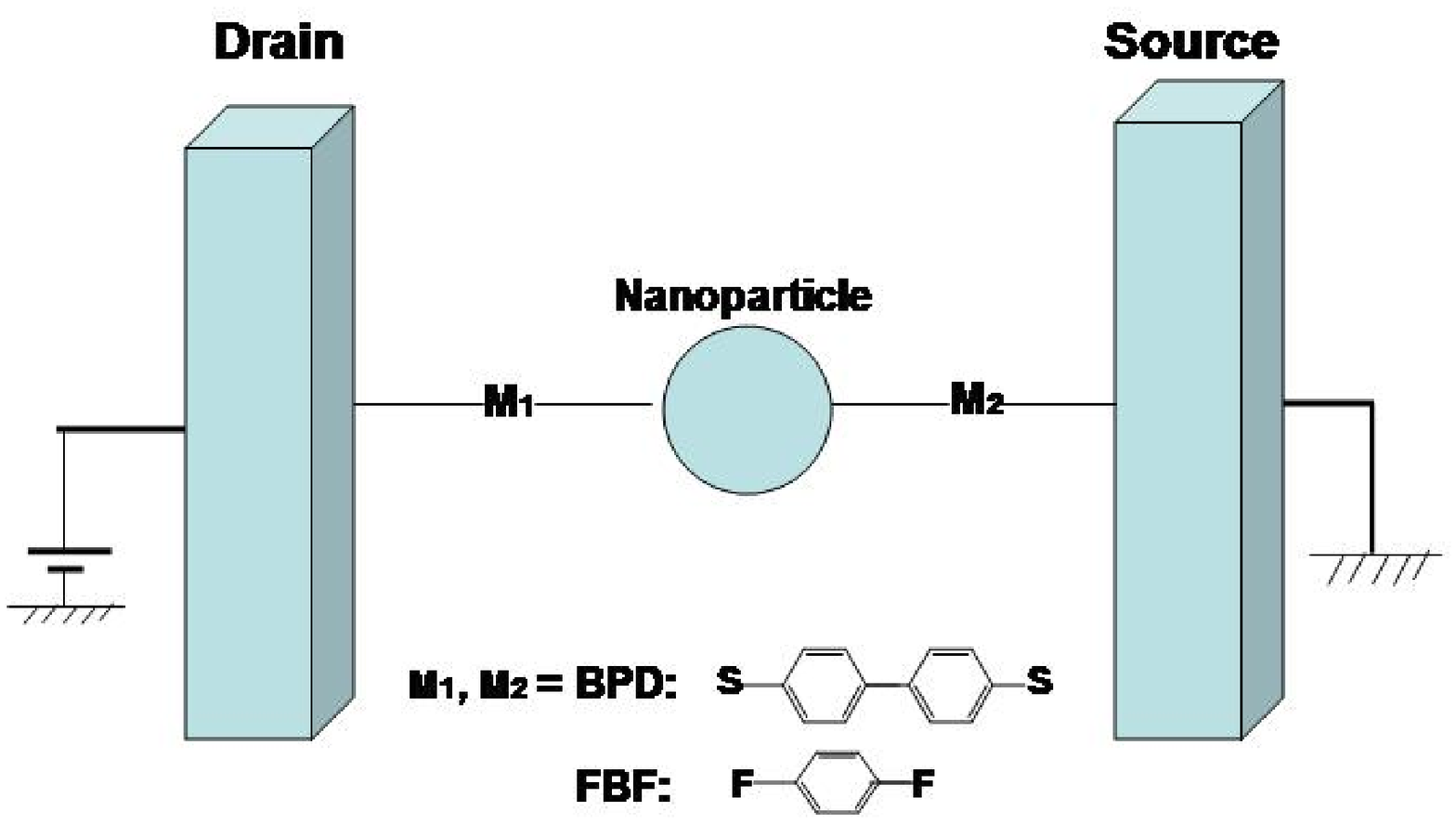}
\caption{\label{xueFig1} Schematic illustration of a metallic nanoparticle 
connected to the source and drain electrodes through two molecular bridges.}
\end{figure}

A schematic illustration of the double-barrier tunneling junction is 
shown in Fig.\ \ref{xueFig1}. 
The system is described by the following Hamiltonian:
\begin{equation}
\label{H1}
H=H_{L}+H_{R}+H_{I}+V+H_{M_{1}}+H_{M_{2}}+H_{T_{1}}+H_{T_{2}}.
\end{equation}
where $H_{\alpha}=
\sum_{ k } \epsilon_{ k \alpha}a_{ k \alpha}^{\dagger}a_{k \alpha} $ and  
$H_{I}=\sum_{l } \epsilon_{ l I}c_{l I}^{\dagger}c_{l I} $
describe the noninteracting electrons in the two leads ($\alpha=L,R$) and 
on the metallic nanoparticle (I) respectively. The indices $k$ and $l$ 
enumerate the electron states of the leads and the nanoparticle. 
The Coulomb repulsion on the nanoparticle $V$ is obtained from 
electrostatic considerations as 
$V(\hat{N})=E_{C}(\hat{N}-n_{x})^{2}$, 
where $E_{C}$ is the charging energy. The background charge $en_{x}$ can be 
separated into an intrinsic background charge $en_{x0}$ existing at zero 
voltage and voltage-dependent polarization charge 
$en_{x}=Q_{P}(V_{L},V_{R})+en_{x0}$. The junction capacitances are 
obtained from $C_{\alpha}=\frac{\partial Q_{P}}{\partial V_{\alpha}}$, where 
$V_{\alpha},C_{\alpha}$ ($\alpha=L,R$) are the voltage and capacitance 
of the two tunnel junctions respectively. The charging energy is 
$E_{C}=e^{2}/2C_{\Sigma}$ and the total capacitance 
$C_{\Sigma}=C_{L}+C_{R}+C_{S}$ is the summation of the two junction 
capacitances and the self-capacitance $C_{S}$ of the nanoparticle. 
Here the background charge $en_{x}$ is induced by the charge transfer 
between the nanoparticle and the linker molecules, whose value at zero 
bias voltage gives the intrinsic background charge $en_{x0}$. 

We assume the molecules can be described by an effective single-particle 
Hamiltonian $H_{M_{i}}=\sum_{m} \epsilon_{im} 
b_{im}^{\dagger}b_{im},i=1,2.$ The transfer 
of single electrons is mediated by tunneling through the molecular bridges 
$H_{T_{i}}=\sum_{mkl} 
 (t_{m k \alpha}b_{im}^{\dagger}a_{k \alpha}
 +t_{m l I}b_{im}^{\dagger}c_{l I}e^{-i \hat{\phi}} + C.C.)$,
where $\alpha$ denotes L (R) for molecule 1 (2).  
The operators $e^{\pm i\hat{\phi}}$ keep track of the change of the charge 
on the metallic nanoparticle by $\pm e$.~\cite{IN92,Schon94,Devoret90} 
The phase operator  
$\hat{\phi}$ is the quantum-mechanical conjugate of the excess electron 
number $\hat{N}$ on the nanoparticle. The charging state $\hat N$ is 
treated independently of the degrees of freedom described by the fermionic 
field operators $c_{lI}^{\dagger},c_{lI}$, which is a good 
approximation for metallic nanoparticles. ~\cite{IN92,Schon94} 
The two electrodes as well as the nanoparticle are treated as large 
equilibrium reservoirs with corresponding Fermi distribution 
$f_{i}(E)=f(E-\mu_{i}),i=L,R,I$. Note that the partition of the voltage 
drop between the two molecular junctions is determined by the 
capacitance ratio $V_{L}/V_{R}=(\mu_{I}-\mu_{L})/(\mu_{R}-\mu_{I})
=C_{R}/C_{L}$ and $eV=\mu_{R}-\mu_{L}$.

Following the standard procedure in NEGF theory of mesoscopic 
transport,~\cite{Wingreen} the current flowing 
through the left contact is given by  
\begin{eqnarray}
I_{L}(t) &=& \frac{ed \hat{N_{L}}}{dt}= 
\frac{2e}{h} Re[\sum_{mkL}t_{mkL}G^{<}_{kLm}(t,t)] 
\nonumber \\
   &=& \frac{2e}{h} \int dt' Tr[G_{M_{1}}^{>}(t,t') \Sigma_{L}^{<}(t',t)
-G_{M_{1}}^{<}(t,t')\Sigma_{L}^{>}(t,t)]
\end{eqnarray} 
where the Green's functions are defined in the standard manner.~\cite{Wingreen} 
The self-energy operator $\Sigma_{L}^{<(>)}$ describes the interaction with 
the left electrode $\Sigma_{L;mn}^{<(>)}(t,t')
=\sum_{k} t_{mkl}(t)g_{L;k}^{<(>)}(t,t')t_{nkL}^{*}(t')$ 
($g_{L}^{<(>)}$ is the corresponding Green's function of the bare left electrode). 
Similar definition applies to the self-energy operators of the right electrode (R) 
and the central island (I). Deriving the equation of motion of the Green's 
function and decoupling the charging dynamics from the single-electron 
dynamics on the metallic nanoparticle 
$\langle c_{l I}^{\dagger}(t)c_{l I}(t') e^{i\hat{\phi} (t')}e^{-i\hat{\phi} (t)} \rangle
=\langle c_{l I}^{\dagger}(t)c_{l I}(t') \rangle 
\langle e^{i\hat{\phi} (t')}e^{-i\hat{\phi} (t)} \rangle $, 
we obtain the current as~\cite{XueMRS}
\begin{eqnarray}
\label{IV}
I_{L} = &-& 2ei \int \frac{dE}{2\pi h} \tilde{T}_{L}(E,V) \int dw  
\nonumber \\
& & [f_{L}(E)(1-f_{I}(E-w))C^{>}(w)+(1-f_{L}(E))f_{I}(E-w)C^{<}(w)],
\end{eqnarray}
where $C^{<(>)}(w)$ are the Fourier transform of the correlation functions 
$C^{<}(t,t')=i \langle e^{i\hat{\phi} (t')}e^{-i\hat{\phi} (t)} \rangle $,
$C^{>}(t,t')=-i \langle e^{-i\hat{\phi} (t)}e^{i\hat{\phi} (t')} \rangle $. 
The ``renormalized'' transmission function is obtained from
$\tilde{T}_{L}(E,V)=
Tr[\Gamma_{L}(E,V) \tilde{G}^{r}_{M_{1}}(E,V) \Gamma_{I}(E,V) 
\tilde{G}^{a}_{M_{1}}(E,V)]$ 
and the ``renormalized'' Green's function is 
$G^{r(a)}_{M1}(E,V)=(E \pm 0^{+}-H_{M1}-\Sigma^{r(a)}_{L}(E,V)-
\int dw \Sigma^{r(a)}_{I}(E-w,V)C^{r(a)}(w))^{-1}$ 
where $C^{r(a)}(w)$ are the Fourier transform of 
$C^{r(a)}(t,t')=\pm \theta (\pm (t-t'))(C^{>}(t,t')-C^{<}(t,t'))$. 

Eq. (\ref{IV}) can be recast into a form suitable for perturbation expansion 
\begin{eqnarray}
\label{IVR}
I_{L} &=& -ie\int dw [\alpha_{L}^{+}(w)C^{>}(w)+\alpha_{L}^{-}(w)C^{<}(w)],
 \nonumber \\
\alpha_{L}^{+}(w) &= & \frac{2}{2\pi h} 
\int dE \tilde{T}_{L}(E,V)f_{L}(E)(1-f_{I}(E-w)), \\
\alpha_{L}^{-}(w) &=& \frac{2}{2\pi h} 
\int dE \tilde{T}_{L}(E,V)(1-f_{L}(E))f_{I}(E-w), \nonumber 
\end{eqnarray}
where $\alpha_{L}^{+(-)}$ is the rate for electron tunneling 
into (out of) the nanoparticle through the left molecular bridge.  
A similar equation can be written down for the current flowing through the 
right contact 
$I_{R} = -ie\int dw [\alpha_{R}^{+}(w)C^{>}(w)+\alpha_{R}^{-}(w)C^{<}(w)]$
where the tunneling rate $\alpha_{R}^{+(-)}$ is evaluated from the 
transmission coefficient through the right molecular junction and 
the corresponding Fermi distributions. 
The condition for current conservation $I_{L}+I_{R}=0$ leads to 
\begin{equation}
\label{IVC}\int dw [\alpha^{+}(w)C^{>}(w)+\alpha^{-}(w)C^{<}(w)]=0, 
\end{equation}
where $\alpha^{+(-)}(w)= \alpha^{+(-)}_{L}(w)+\alpha^{+(-)}_{R}(w)$. 
Eqs. (\ref{IVR}) and (\ref{IVC}) are the central result of this paper which 
separate the description of electron tunneling through the molecular 
bridges from that of the charging dynamics of the nanoparticle through 
the correlation functions $C^{<(>)}$. 

The theory of the correlation functions $C^{<(>)}$ has been developed 
by Sch\"{o}ller,Sch\"{o}n and coworkers in a series of publications based on 
a real-time perturbation theory of the reduced density matrix describing the 
charging state $\hat{N}$ of the metal island, ~\cite{Schon94,Schon98} 
where the perturbation parameter is dimensionless conductance of the 
tunnel junction $\alpha_{0}=h/2e^{2}R_{T}$. In particular, the correlation 
function $C^{<(>)}$ can be expressed as the superposition of 
contributions due to the charging 
state transition from $n$ to $n+1$ as $C^{<(>)}(w)=\sum_{n}C^{<(>)}(w,n)$. 
The current conservation condition (\ref{IVC}) can also be shown to hold 
for each part $C^{<(>)}(w,n)$,~\cite{Schon98}  
$\int dw [\alpha^{+}(w)C^{>}(w,n)+\alpha^{-}(w)C^{<}(w,n)]=0$. 
In the lowest order perturbation theory with respect to $\alpha_{0}$, 
$C^{<(>)}(w,n)$ are related to the probability 
$P_{n}$ of finding $n$ excess electrons as~\cite{Schon98} 
$C^{<}(w,n)=2\pi i P_{n+1}\delta (w-\Delta_{n})$ and  
$C^{>}(w,n)=-2\pi i P_{n}\delta (w-\Delta_{n})$. 
Here $\Delta_{n}$ is the energy difference of the adjacent charging states 
$\Delta_{n}=V(n+1)-V(n)=E_{C}[1+2(n-n_{x})]$. 
Substituting the above formulas into Eqs.\ (\ref{IVR}) and (\ref{IVC}), we 
obtain a set of rate equations formally equivalent to the ``orthodox'' 
theory of single-electron tunneling 
\begin{equation}
\label{rate}
\alpha^{+}(\Delta_{n})P_{n}-\alpha^{-}(\Delta_{n})P_{n+1}=0, 
\end{equation}
Compared to the ``orthodox'' theory,  
the above theory allows us to take full account of the microscopic description 
of the electronic processes in the molecular junction. In particular, 
considerable simplification can be achieved in the ``wide-band limit'' of 
the electrodes, i.e., if the surface density of states of the electrodes 
is approximately constant in the energy interval of $\Delta_{n}$. 
The ``renormalization'' of the nanoparticle self-energy and transmission 
coefficient can then be neglected, which is true for common simple and noble 
metals and for nanoparticle size sufficiently large that energy quantization can be 
neglected.~\cite{Note1}   

Here we apply the theory to single-electron tunneling through a gold 
nanoparticle connected to two gold electrodes using biphenyl dithiolate 
(BPD) and difluorobenzene (FBF) molecules. 
Transport through gold-BPD-gold and gold-FBF-gold junctions  
have been recently studied in detail using a first-principles self-consistent 
matrix Green's function theory of electron transport in molecular 
junctions,~\cite{Xue01,Xue03} from which we obtain the voltage-dependent 
background charge $en_{x}$ induced by the charge transfer from the 
molecule to the nanoparticle and the transmission coefficient for electron 
tunneling through the metal-molecule-nanoparticle junction.~\cite{Note} 
The charge transfer at zero bias gives the intrinsic background 
charge $en_{x0}$, while its derivative with respect to bias voltage gives 
the capacitance of the molecular junction.~\cite{Note2} The calculated 
dimensionless conductance, capacitance and the intrinsic background charge 
for BPD (FBF) molecular junction are $0.0183$ ($0.0023$), 
$0.037$ ($0.043$) aF and $-0.623$ ($-0.805$) number of electrons 
respectively. Note the total intrinsic background charge on the nanoparticle 
is the sum of contributions from the two molecular junctions. Despite the 
shorter length of the FBF molecule, the resistance of the FBF junction is 
much higher than that of the BPD molecule due to the less favorable 
energy-level lineup relative to the gold Fermi-level.~\cite{Xue01,Xue03} 

Given the transmission coefficient and capacitance of the molecular junction, 
the rate equation (\ref{rate}) is solved using the standard 
procedure,~\cite{Amman91} from which we calculate the terminal current as  
\begin{equation} 
I_{L(R)}=2\pi e \sum_{n} P_{n}
[\alpha^{+}_{L(R)}(\Delta_{n})-\alpha^{-}_{L(R)}(\Delta_{n-1})]. 
\end{equation}
The average excess charge $e \langle \hat{N} \rangle=e\sum_{N}NP_{N}$ 
gives the net charge transferred onto the nanoparticle due to the discrete  
electron tunneling across the molecular bridges, while the average charge 
$e (\langle \hat{N} \rangle-n_{x})$  gives the charging configuration 
which determines the electrostatic energy cost for charging the nanoparticle.  

\begin{figure}
\includegraphics[height=3.0in,width=3.0in]{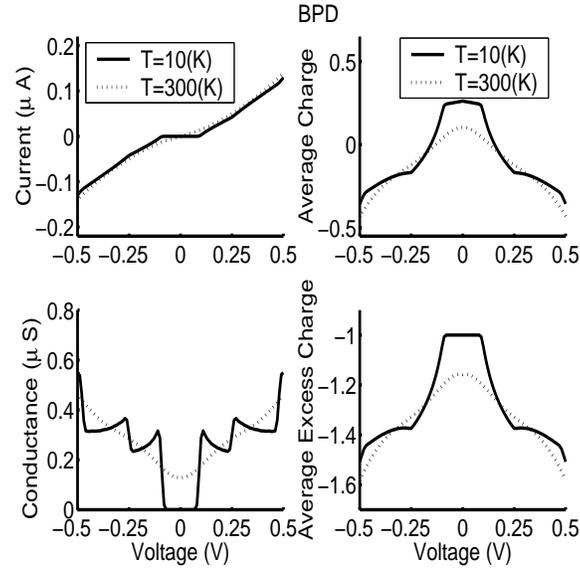}
\caption{\label{xueFig2} Single-electron tunneling through 
a $10$ nm-diameter gold nanoparticle connected to the source and 
drain electrodes through the BPD molecule. Left figure shows the 
current-voltage and conductance-voltage characteristics. Right figure shows 
the average charge $e(\hat{N}-n_{x})$ and average excess charge $e\hat{N}$  
on the nanoparticle in unit of electron charge $e$. }
\end{figure}

\begin{figure}
\includegraphics[height=3.0in,width=3.0in]{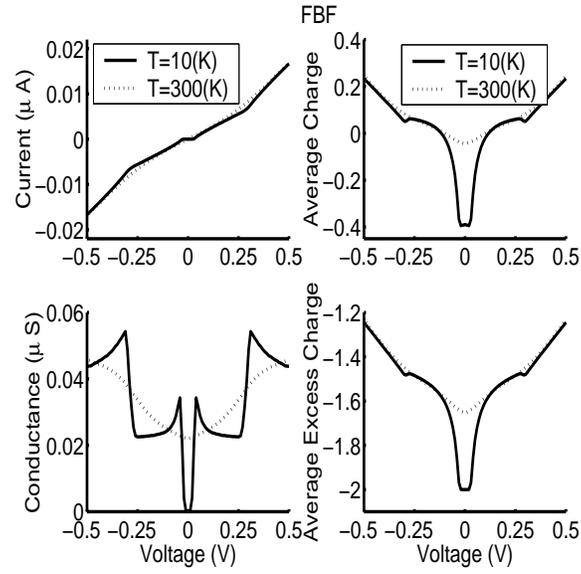}
\caption{\label{xueFig3} Single-electron tunneling through 
a $10$ nm-diameter gold nanoparticle connected to the source and 
drain electrodes through the FBF molecule as in Fig. \ref{xueFig2}.}
\end{figure}

Figs. \ref{xueFig2} and \ref{xueFig3} show the calculated results 
for single-electron tunneling through a $10$ nm-diameter gold 
nanoparticle connected to gold electrodes through two 
BPD or FBF bridges at temperatures of $10(K)$ and $300(K)$ respectively. 
The self-capacitance of the nanoparticle is evaluated using the electrostatic 
formula for a conducting sphere in front of a conducting 
plane,~\cite{SETBook} which gives a charging energy of $E_{C}=88(meV)$. 
The intrinsic background charge $n_{x0}$ on the central nanoparticle is 
closer to half integer in the case of the FBF molecule ($-1.61$) than that 
of the BPD molecule ($-1.25$), so the Coulomb gap in the 
current/conductance-voltage characteristics is smaller. Note at low 
temperature the average excess electron $\langle \hat{N} \rangle $ on the 
nanoparticle at low bias equals the nearest integer to $n_{x0}$. Since the 
background charge $n_{x}$ varies slowly with the applied bias voltage within 
the voltage range studied, both the average charge and the average excess charge 
show similar voltage dependence. The peaks in the conductance-voltage 
characteristics correlate with the change in the average excess electron on the 
nanoparticle at low temperature.  

\begin{figure}
\includegraphics[height=3.0in,width=3.0in]{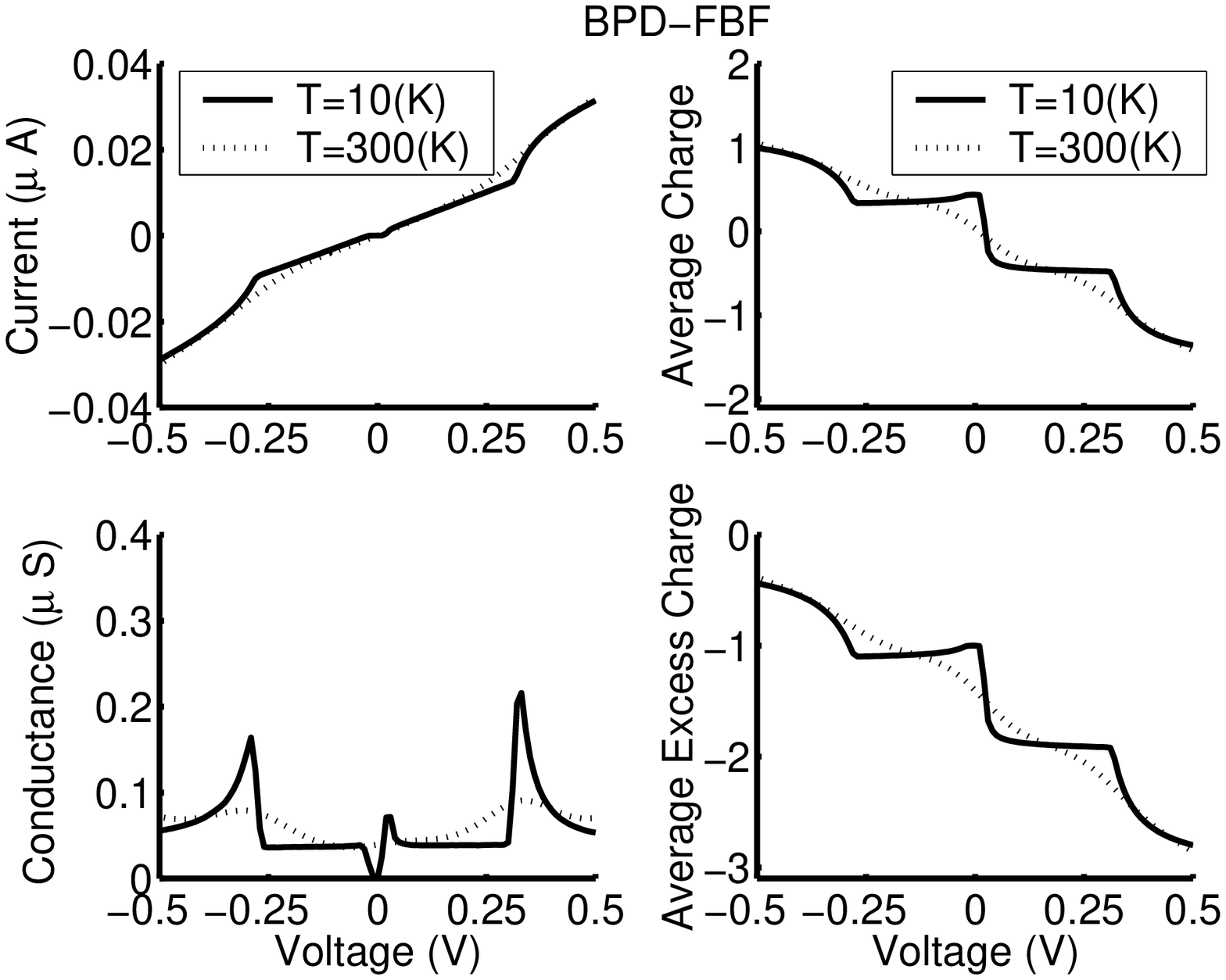}
\caption{\label{xueFig4} Single-electron tunneling through 
a $10$ nm-diameter gold nanopartcile connected to the source 
electrode through the BPD molecule and the drain electrode through 
the FBF molecule as in Fig. \ref{xueFig2}.}
\end{figure}

Fig. \ref{xueFig4} shows the calculated results for single-electron tunneling 
through a $10$ nm-diameter gold nanoparticle connected to left gold electrode 
through BPD molecule and connected to right gold electrode through FBF 
molecule. Since $R_{FBF}/R_{BPD} \gg 1$, varying bias voltage leads to 
stepwise charging of the central nanoparticle at low temperature. Here 
the intrinsic background charge $n_{x0}\approx -1.43$ and 
$C_{FBF}/C_{BPD}>1$, 
the shape of the Coulomb staircase corresponds to case $\bf{I}$ as 
discussed by Hanna and Tinkham~\cite{Tinkham91} and is well understood. 
The small Coulomb gap is again due to the fact that $n_{x0}$ is close to 
half integer.  

To conclude, we have presented a microscopic theory of single-electron 
tunneling through molecular-assembled metallic nanoparticles, which 
combines the theory of transport through a molecular junction 
with the description of the charging dynamics on the nanoparticle. This 
allows us to take full account of the microscopic description of the 
electronic processes in the metal-molecule-nanoparticle junction 
as well as the strong Coulomb interaction on the nanoparticle.

This work was supported by the DARPA Molectronics program, 
the DoD-DURINT program and the NSF Nanotechnology Initiative.

\newpage
 
\begin{references}
\bibitem[*]{Xue} Corresponding author. Email: ayxue@chem.nwu.edu 
\bibitem{Heath98} C.P. Collier, T. Vossmeyer and J.R. Heath, Annu. Rev. 
Phys. Chem.\ {\bf 49}, 371 (1998); C.B. Murray, C.R. Kagan and M.G. Bawendi, 
Annu. Rev. Mater. Sci. {\bf 30}, 545 (2000).
\bibitem{Bio} J.J. Storhoff and C.A. Mirkin, Chem. Rev. {\bf 99}, 1849 
(1999). 
\bibitem{Andres96} R.P. Andres, T. Bein, M. Dorogi, S. Feng, J.I. Henderson, 
C.P. Kubiak, W. Mahoney, R.G. Osifchin and R. Reifenberger, Science 
{\bf 272}, 1323 (1996). 
\bibitem{SE} S.H.M. Persson, L. Olofsson and L. Gunnarsson, 
Appl.\ Phys.\ Lett.\ {\bf 74}, 2546 (1999); J.R. Petta, D.G. Salinas 
and D.C. Ralph, \emph{ibid.} {\bf 77}, 4419 (2000); T. Ohgi, H.-Y. Sheng, 
Z.-C. Dong, H. Nejoh and D. Fujita, \emph{ibid.} {\bf 79}, 2453 (2001); 
C.A. Berven, M.N. Wybourne, L. Clarke, L. Longstreth, J.E. Hutchinson 
and J.L. Mooster, J. Appl. Phys.\ {\bf 92}, 4513 (2002); K.-H. M\"{u}ller, 
J. Herrmann, B. Raguse, G. Baxter and T. Reda, Phys. Rev. B {\bf 66}, 
75417 (2002).      
\bibitem{IN92} D.V. Averin and K.K. Likharev, in \emph{Mesocopic Phenomena 
in Solids}, edited by B.L. Altshuler, P.A. Lee and R.A. Webb (Elsevier, 
Amsterdam, 1991); G.-L. Ingold and Yu.V. Nazarov, in \emph{Single Charge 
Tunneling}, edited by H. Grabert and M.H. Devoret (Plenum, New York, 1992).
\bibitem{JGA00} C. Joachim, J. K. Gimzewski and A. Aviram, Nature 
{\bf 408}, 541 (2000); A. Nitzan and M.A. Ratner, Science {\bf 300}, 
1384 (2003).  
\bibitem{Xue01} Y. Xue, S. Datta and M.A. Ratner, J. Chem. Phys. {\bf 115}, 
4292 (2001); Chem. Phys. {\bf 281}, 151 (2002).
\bibitem{Xue03} Y. Xue and M.A. Ratner, Phys. Rev. B {\bf 68}, 115406 
(2003); to be published.
\bibitem{Buttiker93} M. B\"{u}ttiker, J. Phys.: Condens. Matter {\bf 5}, 
9379 (1993).  
\bibitem{Tinkham91} A.E. Hanna and M. Tinkham, Phys. Rev. B {\bf 44}, 5919 
(1991). 
\bibitem{Schon94} H. Schoeller and G. Sch\"{o}n, Phys. Rev. B {\bf 50}, 
18436 (1994); Physica {\bf 203}, 423 (1994).  
\bibitem{Schon98} J. K\"{o}nig, H. Schoeller and G. Sch\"{o}n, Phys.\ Rev. 
Lett.\ {\bf 78}, 4482 (1997); Phys. Rev. B {\bf 58}, 7882 (1998).   
\bibitem{Devoret90} M.H. Devoret, D. Esteve, H. Grabert, G.-L. Ingold, 
H. Pothier and C. Urbina, Phys.\ Rev.\ Lett.\ {64}, 1824 (1990).
\bibitem{Wingreen} Y. Meir and N.S. Wingreen, Phys. Rev. Lett.\ {\bf 68}, 
2512 (1992); H. Haug and A-P. Jauho, \emph{Quantum Kinetics in 
Transport and Optics of Semiconductors} (Springer-Verlag, Berlin, 1996).  
\bibitem{XueMRS} Y. Xue and M.A. Ratner, Materials Research Society 
Proceedings {\bf 735}, C5.5 (2003). 
\bibitem{Note1} Depending on the molecules chosen, the molecular junction 
may become sufficiently conducting such that higher order effects including 
quantum fluctuation of charges on the nanoparticle and cotunneling effects 
become important. The microscopic theory presented here provides a 
systematic way of investigating such effects following the procedures described 
in Refs.\ \onlinecite{Schon94} and \onlinecite{Schon98}. But for 
single-electron device applications, molecular junction with high resistance is 
desirable and the lowest-order perturbation theory is sufficient. The choice of 
the molecules studied here reflects this consideration.  
\bibitem{Note} Since the metal-molecule interaction is a local phenomenon, 
we neglect the curvature of the nanoparticle surface and calculate 
the electronic structure of metal-molecule-nanoparticle junction in 
the same way as that of the metal-molecule-metal junction in 
Refs.\ \onlinecite{Xue01} and \onlinecite{Xue03}.   
\bibitem{Note2} The self-consistent calculation is performed in a ``extended 
molecule'' region including both the molecule and the perturbed surface atoms 
of the electrodes and the nanoparticle. The transferred charge and therefore 
the background charge $n_{x}$ is obtained by integrating the (self-consistent) 
electron density distribution over the region occupied by the perturbed surface 
atoms of the nanoparticle.   
\bibitem{Amman91} M. Amman, R. Wilkins, E. Ben-Jacob, P.D. Maker 
and R.C. Jaklevic, Phys. Rev. B {\bf 43}, 1146 (1991).  
\bibitem{SETBook} C. Wasshuber, \emph{Computational Single-Electronics} 
(Springer-Verlag, Wien, 2001). 
\end{references}
\end{document}